\documentclass[fleqn,usenatbib]{mnras}

\usepackage{newtxtext,newtxmath}

\usepackage[T1]{fontenc}

\DeclareRobustCommand{\VAN}[3]{#2}
\let\VANthebibliography\thebibliography
\def\thebibliography{\DeclareRobustCommand{\VAN}[3]{##3}\VANthebibliography}

\usepackage{graphicx}	
\usepackage{amsmath}	
\usepackage{physics}
\usepackage{comment}

\newcommand{\modA}[1]{$\Lambda$CDM-$\delta$}
\newcommand{\modB}[1]{PBH-$\delta$}
\newcommand{\modC}[1]{PBH-lognormal}

\defcitealias{Ziparo+2022}{Z22}
\defcitealias{Hasinger2020}{H20}

\title[PBHs as NIRB sources]{Primordial Black Holes as Near Infrared Background sources}

\author[D. Manzoni et al.]{
D. Manzoni,$^{1}$\thanks{E-mail: daniele.manzoni@sns.it}
F. Ziparo,$^{1}$
S. Gallerani,$^{1}$
A. Ferrara$^{1}$
\\
$^{1}$Scuola Normale Superiore, Piazza dei Cavalieri 7, I-56126 Pisa, Italy
}

\date{Accepted XXX. Received YYY; in original form ZZZ}

\pubyear{2023}

\begin{document}
\label{firstpage}
\pagerange{\pageref{firstpage}--\pageref{lastpage}}
\maketitle

\begin{abstract}
The near infrared background (NIRB) is the collective light from unresolved sources observed in the band 1--10 $\micron$. The measured NIRB angular power spectrum on angular scales $\theta \gtrsim 1$ arcmin  exceeds by roughly two order of magnitudes predictions from known galaxy populations. The nature of the sources producing these fluctuations is still unknown. Here we test primordial black holes (PBHs) as sources of the NIRB excess. Considering PBHs as a cold dark matter (DM) component, we model the emission of gas accreting onto PBHs in a cosmological framework. We account for both accretion in the intergalactic medium (IGM) and in DM haloes. We self consistently derive the IGM temperature evolution, considering ionization and heating due to X-ray emission from PBHs. Besides $\Lambda$CDM, we consider a model that accounts for the modification of the linear matter power spectrum due to the presence of PBHs; we also explore two PBH mass distributions, i.e. a $\delta$-function and a lognormal distribution. For each model, we compute the mean intensity and the angular power spectrum of the NIRB produced by PBHs with mass 1--$10^3~\rm M_{\sun}$. In the limiting case in which the entirety of DM is made of PBHs, the PBH emission contributes  $<1$ per cent to the observed NIRB fluctuations. This value decreases to $<0.1$ per cent if current constraints on the abundance of PBHs are taken into account. We conclude that PBHs are ruled out as substantial contributors to the NIRB. 
\end{abstract}




\section{Introduction}
The Near Infrared Background (NIRB) is the diffuse radiation of cosmological origin observed after subtracting the local foregrounds in the band 1--10 $\micron$ \citep{LibraeReview}. Since early studies by \citet{Partridge1967}, the NIRB has been considered a valuable tool to investigate the emission from the first stars and galaxy populations, as ultraviolet (UV) and optical light from high-z sources is redshifted to the near-infrared band. 

Actual measurements of the mean NIRB intensity \citep{Tsumura+2013,Matsumoto+2015,Sano+2015,Matsuura+2017} give a lower bound $I\gtrsim 10~\text{nW m}^{-2} \text{sr}^{-1}$, in excess with respect to the contribution of known galaxy populations derived from galaxy number counts \citep{Driver+2016}. However, direct measurements of the NIRB suffer from large uncertainties due to the subtraction of foregrounds \citep{Leinert+1998}, namely interplanetary dust emission (zodiacal light), galactic stars light and galactic interstellar medium radiation (cirrus). 

Being foregrounds smooth, a more robust technique is computing the power spectrum of NIRB fluctuations \citep{Kashlinsky+1996,Kashlinsky+2000}, to which foregrounds contribution is limited. Moreover, from the power spectrum measurements, a lower limit to the $I_\nu$ contribution from unknown sources can be derived \citep{Kashlinsky+2007}. The latest measurements of the NIRB power spectrum \citep{Cooray+2012,Kashlinsky+2012} established an excess power on scales larger then $\gtrsim 1$ arcmin, irreconcilable with emission from known galaxies up to $z \sim 5$ \citep{Helgason+2012}. The origin of such a signal is still unknown. 

Population III stars (PopIII) were one of the first hypothesis proposed about the sources of the NIRB excess \citep{Santos+2002,Salvaterra+2003}. Although intriguing, this idea was soon after discarded because of the very high formation efficiency required \citep{Madau+2005} and since it would overpredict the number of high-z dropout galaxies \citep{Salvaterra+2006}. Several works explored the possibility of high redshift galaxies ($z \gtrsim 5$) being the sources of the NIRB excess, but models failed to reproduce the required levels of fluctuations \citep{Fernandez+2010,Cooray+2012_EoR,Yue+2013,Helgason+2016}. 

An alternative solution was the intrahalo light (IHL), i.e light from stars stripped from their parent galaxy \citep{Cooray+2012,Cheng+2022}. Despite its success in reproducing observations, such idea has to rely on poorly understood abundance of intrahalo stars \citep{Ferrara2012}. Moreover, this model cannot account for the observed cross-correlation of the NIRB with the soft-X background (SXB) \citep{Cappelluti+2013,Cappelluti+2017}. Such a feature is difficult to explain even with galaxies spectra, but could be naturally justified by X-ray emission from accretion disks around black holes. \citet{Yue+2013_DCBH} developed a model to explain both NIRB fluctuations and NIRB-SXB cross correlation with accreting direct collapse black holes (DCBHs), even though it is unclear whether the specific conditions of DCBHs formation are actually realized during cosmic evolution \citep{Latif+2016}.

Given the puzzling nature of the NIRB excess, a new scenario has been recently suggested, invoking Primordial Black Holes (PBHs) \citep{Kashlinsky2016,Cappelluti+2022}. PBHs are black holes formed deep into radiation dominated era from the collapse of overdensity peaks \citep{Zeldovich+1967,Carr+1974} and interest on them have been rejuvenated after the first detection of gravitational waves from black holes merger \citep{Abbott+2016,Bird+2016,Blinnikov+2016,Sasaki+2016}. The primordial origin of LIGO/VIRGO black holes is a viable solution to explain their observed mass spectrum and merger rates \citep{Raidal+2017,AliHaimoud+2107_merger_rates,Wong+2021}. Moreover, they could justify why most of the measured effective spins are close to zero \citep{Abbott+2019_properties,DeLuca+2020} and could accomodate for black holes with masses in the pair-instability supernovae mass gap (45--120~$\rm M_{\sun}$) \citep{Abbott+2020_PISN,DeLuca+2021,OBrien+2021} and in the low mass gap (2.5--5~$\rm M_{\sun}$) \citep{Abbott+2020_low_gap1,Abbott+2020_low_mass_gap2,Clesse+2022}. Finally, the recent evidence of a gravitational-wave background reported by the NANOGrav collaboration \citep{Agazie+2023} could directly probe PBH formation from high amplitude peaks of the primordial power spectrum \citep{Clesse+2017, Ville+2021,Franciolini+2023}.

A key aspect about PBHs is that they were proposed as cold dark matter candidates \citep{Chapline1975}. This hypothesis has been investigated in a plethora of studies, providing constraints on the fraction of DM comprised by PBHs (see \citet{Carr+2020} for a review). The presence of PBHs would entail a variety of astrophysical phenomena, such as gamma rays emission from evaporating PBHs \citep{Laha2019,Coogan+2020}, microlensing effects \citep{Nikura+2019,Blaineau+2022} and disruption of wide binaries or ultra-faint dwarfs \citep{Monroy+2014,Brandt2016}. In addition, accreting PBHs would impact the CMB spectrum and anisotropies \citep{Poulin+2017,Serpico+2020}, the 21 cm power spectrum \citep{Mena+2019} and would produce radio and X-ray backgrounds \citep{Cappelluti+2022,Ziparo+2022}. 

When deriving constraints on the abundance of PBHs, it is commonly assumed that PBHs have the same mass (i.e. a $\delta$-function), although these constraints actually depend on the adopted PBH mass function \citep{Kuhnel+2017}. In particular, PBH formation models in slow-roll inflation predict an approximately lognormal mass function \citep{Dolgov+1993,Kannike+2017}, while latest simulations of PBH formation across the QCD epoch derived a mass function peaked around $M_{\text{PBH}} \sim 1~\rm M_{\sun}$, with a non-trivial shape departing from lognormal \citep{Franciolini+2022}.

If PBHs constitute a fraction of dark matter, they would add a poissonian component to the matter power spectrum \citep{Meszaros1975,Afshordi+2003,Alihaimoud2018}, accelerating structure formation and consequently enhancing the abundance of haloes in which stars can form \citep{Kashlinsky2016}. This effect on the star formation process is particularly relevant for what concerns the NIRB excess puzzle, since a higher star formation rate density at high-z can then provide the required levels of NIRB fluctuations \citep{Cappelluti+2022}. Moreover, PBHs could directly contribute to the NIRB with the radiation emitted by accreting gas from their surroundings. 

\citet[hereafter \citetalias{Hasinger2020}]{Hasinger2020} computed cosmic backgrounds from gas accretion onto PBHs and could recover only 0.3 per thousand of the NIRB with his model. However, \citetalias{Hasinger2020} considered gas accretion only in the intergalactic medium (IGM), while PBHs could accrete matter also in dense virialized structures, i.e DM haloes. In particular, \citet[hereafter \citetalias{Ziparo+2022}]{Ziparo+2022} have shown that the contribution of PBHs accreting in DM haloes to X-ray and Radio backgrounds is $> 60$ per cent larger than those accreting in the IGM.

In this paper, following the model by \citetalias{Ziparo+2022}, we compute the NIRB produced by PBHs taking into account both PBH accretion in DM haloes and a self-consistent treatment of X-ray ionization and heating of the IGM. We further improve the \citetalias{Ziparo+2022} model both considering the modification of the matter power spectrum induced by the presence of the PBHs, previously neglected, and generalizing the framework to extended mass functions. In Section~\ref{sec:Methods} we summarize the basic model and present its extensions. In Section~\ref{sec:Results} we present the main results of this work. 
Finally, we state our conclusions in Section~\ref{sec:Conclusion}. 

Throughout the paper we assume a flat Universe with the following cosmological parameters: $\Omega_m = 0.3075$, $\Omega_{\Lambda} = 1 - \Omega_m$, $\Omega_b = 0.0486$, $H_0 = 67.74~ \text{km s}^{-1}\text{Mpc}^{-1}$ , $n_s = 0.965$ and $\sigma_8 = 0.811$ \citep{Planck2015}.

\section{Methods }
\label{sec:Methods}
To investigate the contribution of PBHs to the NIRB we rely on the formalism described in \citetalias{Ziparo+2022}. We first revisit their model in order to introduce the framework (Section \ref{subsec:Zip_cosmological_distribution}, \ref{subsec:Zip_accretion}). We then compute the intensity and angular power spectrum of the NIRB in Section~\ref{subsec:NIRB}. In the last two Sections we extend the model to account for the modification of matter power spectrum induced by PBHs (Section \ref{subsec:PBH-LCDM} ) and extended mass functions (Section \ref{subsec:ext_mass_func}). 

\subsection{Cosmological  distribution of PBHs}
\label{subsec:Zip_cosmological_distribution}
Assume that a DM fraction $f_{\text{PBH}}$ is made of PBHs of mass $M_{\text{PBH}}$. DM distribution on cosmological scales can be described as a diffuse component with density equal to the mean DM density, and virialized regions where matter has collapsed into DM haloes. As PBHs are distributed as the DM, we decompose the number density of PBHs as 
\begin{equation}
    n_{\text{PBH}}(z) = \frac{f_{\text{PBH}}\Omega_{\text{DM}}\rho_c (1+z)^3}{M_{\text{PBH}}} = n_{\text{PBH}}^{\text{IGM}}(z)+n_{\text{PBH}}^h(z),
    \label{eq:n_PBH}
\end{equation}
where $n_{\text{PBH}}^{\text{IGM}}$ ($n_{\text{PBH}}^h$) is the number density of PBHs in the intergalactic medium (haloes). The abundance of PBHs in haloes is related to the collapsed fraction of DM in haloes $f_{\text{coll}}$, which can be computed as
\begin{equation}
    f_{\text{coll}}(M_h,z) = \text{erfc}\Big(\frac{\delta_{\text{crit}}(z)}{\sigma_M}\Big)
    \label{eq:f_coll}
\end{equation}
in the Press-Schechter formalism \citep{PressSchechter}. Here $\delta_{\text{crit}}(z) = 1.68/D(z)$ is the critical overdensity for collapse, $D(z)$ is the growth factor and $\sigma_M^2$ is the mass variance of the linearly extrapolated matter overdensity field. Thus, the number density of PBHs in the IGM and in haloes are
\begin{equation}
    n_{\text{PBH}}^{\text{IGM}}(z) = (1-f_{\text{coll}})n_{\text{PBH}}(z), \quad n_{\text{PBH}}^{h}(z) = f_{    \text{coll}}n_{\text{PBH}}(z).
    \label{eq:n_PBH_IGM_HALO}
\end{equation}
\subsubsection{PBH distribution inside haloes}

The distribution of PBHs inside haloes follows the DM density profile,  here assumed to be NFW \citep[][]{NFW1996}:
\begin{equation}
    \rho_{\text{DM}}(x) = \frac{\rho_c \delta_{c}}{cx(1+cx)^2},
    \label{eq:NFW_profile}
\end{equation}
where $x = r/r_{\text{vir}}$ is the radial distance in virial radius units and $c$ is the concentration parameter from \citet{Maccio+2007}. Following \citetalias{Ziparo+2022}, we model its redshift evolution as $c \propto (1+z)^{-1}$. 
The parameter $\delta_c$ is a function of both the concentration parameter and the overdensity at the collapse redshift $\Delta_c$ \citep{Barkana+2001}:
    \begin{equation}
    \delta_c = \frac{\Delta_c}{3}\frac{c^3}{\mathrm{ln}(1+c)-c/(1+c)},
    \label{eq:delta_c}
\end{equation}
with $\Delta_c = 18\pi^2+82d-39d^2$, $d = \Omega_m^z-1$ and $\Omega_m^z = \Omega_m(1+z^3)/(\Omega_m(1+z^3)+\Omega_{\Lambda})$. Being PBHs distributed as DM, the number of PBHs within radius $r$ an $r+\dd r$ is
\begin{equation}
    \dd N_{\text{PBH}}(r) = \frac{f_{\text{PBH}}}{M_{\text{PBH}}}4\pi r^2\rho_{\text{DM}}(r)\dd r.
    \label{eq:N_PBH}
\end{equation}

\subsection{PBHs accretion}
\label{subsec:Zip_accretion}
To estimate the accretion rate of gas onto PBHs, we adopt the Bondi--Hoyle--Lyttleton formula \citep{Bondi1952,Edgar2004}: 
 \begin{equation}
    \dot{M} = \lambda 4\pi \frac{G^2M_{\text{PBH}}^2\rho_b}{(c_s^2+v_{\text{BH}}^2)^{3/2}},
    \label{eq:BHL_accretion_rate}
\end{equation}
where $\rho_b$ and $c_s$ are the density and sound speed of the accreting gas, respectively, $v_{\text{BH}}$ is the relative velocity between the PBH and the gas, and $\lambda$ is the accretion parameter that accounts for non gravitational effects (i.e radiative feedback, gas pressure, outflows). Following \citet{Poulin+2017}, we adopt the value $\lambda = 0.01$, which is a benchmark for an advection dominated accretion flow \citep[ADAF,][]{Yuan+2014}. 

Accretion conditions in the IGM and inside haloes differ substantially: in the following we describe the relevant physical quantities, i.e. $\rho_b, c_s$ and $v_{\text{BH}}$, separately for the two cases. 
\subsubsection{Accretion in the IGM}
Following \citet{Ricotti+2008}, we assume a uniform gas density in the IGM, equal to
\begin{equation}
    \rho_{\text{IGM}}(z) = 250\,\mu m_p \Big(\frac{1+z}{1000}\Big)^{3}\text{g cm}^{-3},
    \label{eq:rho_IGM}
\end{equation}
where $\mu = 1.22$ is the mean molecular weight for a gas of primordial composition and $m_p$ is the proton mass. The sound speed of the gas is given by:
\begin{equation}
    c_s = \sqrt{\frac{k_{B} T_{\text{IGM}}}{\mu m_p}},
    \label{eq:sound_speed_IGM}
\end{equation}
where $k_B$ is the Boltzmann constant and $T_{\text{IGM}}$ is the IGM temperature. The relative velocity between baryons and PBHs is gaussianly distributed on linear scales, hence its modulus follows a maxwellian distribution, with variance given by \citep{Alihaimoud+2017}
\begin{equation}
    \sigma^2_{\text{rel}}(z) \equiv \langle v_{\text{BH}}^2 \rangle = 30\;min\Big[1,(1+z)/1000 \Big] \text{km s}^{-1}.
\end{equation}
To properly account for the distribution of relative velocities, it is useful to define an effective velocity $v_{\text{eff}}$ \citep{Ricotti+2008}, whose analytical expression is
\citep{Mena+2019}
\begin{equation}
    v_{\text{eff}} = \sigma_{\text{rel}}(z)\Big[\Big( \frac{3}{2}\Big)^{3/2}U\Big(\frac{3}{2},1,\frac{3}{2}\Big(\frac{\sigma_{\text{rel}}}{c_s}\Big)^{-2}\Big)\Big]^{-1/3},
    \label{eq:v_eff}
\end{equation}
where $U(a,b,z)$ is the confluent hypergeometric function of second kind. 
The accretion rate of PBHs in the IGM is finally obtained by substituting the relevant quantities computed above in equation~(\ref{eq:BHL_accretion_rate}).
\subsubsection{Accretion within haloes}
To model the internal  structure of haloes, we assume that the gas is in thermal equilibrium at the virial temperature $T_{\text{vir}}$. Moreover, we impose hydrostatic equilibrium between DM and gas. Given these assumptions, the density profile of gas is described by the following equation \citep{Makino+1998}:
\begin{equation}
    \rho_b(r) = \rho_{b,0}\exp \Big[-\frac{\mu m_p}{2k_B T_{\text{vir}}}\Big(V_{\text{esc}}^2(0)-V_{\text{esc}}^2(r) \Big)\Big],
    \label{eq:baryon_profile}
\end{equation}
where $V_{\text{esc}}$ is the escape velocity, given by:
\begin{equation}
    V_{\text{esc}}^2(r) = 2\int_r^{r_{ \text{vir}}} \dd r'\frac{GM(r')}{r'^2},
\end{equation}
and $\rho_{b,0}$ is a normalization constant set by imposing:
\begin{equation}
    4\pi \int_{0}^{r_{\text{vir}}} \dd r r^2\rho_b(r) = \frac{\Omega_{b}}{\Omega_{\text{DM}}}M_h,
    \label{eq:baryon_profile_normalization}
\end{equation}
where $M_h$ is the halo mass, $\Omega_{b}$ and $\Omega_{\text{DM}}$ are the total baryon and DM densities in units of the critical density. The sound speed in haloes can be computed via equation~(\ref{eq:sound_speed_IGM}), substituting $T_{\text{IGM}}$ with $T_{\text{vir}}$. As a consequence of hydrostatic equilibrium assumption, we set $v_{\text{BH}} = 0$.   

\subsection{NIRB}
\label{subsec:NIRB}
To compute the specific luminosity of PBHs we follow \citetalias{Ziparo+2022}. Given the accretion rate $\dot{M}$, the bolometric luminosity of a single PBH is $L = \varepsilon\dot{M}c^2$, where $\varepsilon = 0.1$ is the radiative efficiency. We assume that, as for astrophysical black holes, the spectrum of PBHs can be described by a double power-law with an exponential cut-off \citepalias{Hasinger2020}:
\begin{equation}
    L_{\nu} \propto \begin{cases}
            \Big(\frac{\nu}{\nu_c}\Big)^{\alpha_{\text{sync}}} & \nu \leq \nu_c \\
            \Big(\frac{\nu}{\nu_c}\Big)^{\alpha} & \nu > \nu_c
    \end{cases}
    \label{eq:spectrum}
\end{equation}
where the cut-off frequency is $\nu_{\text{cut}} = 200$ keV and $\alpha = -0.7$. Below the critical frequency $\nu_c = \lambda_c/c$, with $\lambda_c = 0.45(M_{\text{PBH}}/ \rm M_{\sun})^{0.4} \micron$, synchrotron emission dominates and the power law index is $\alpha_{\text{sync}} = 1.86$ \citepalias{Hasinger2020}. The above spectral shape is consistent with an ADAF accretion model with accretion rates $\dot{m} = \dot{M}/\dot{M}_{\text{EDD}} \gtrsim 10^{-2}$, which holds for those PBHs producing the bulk of the background radiation in our model. We fix the normalization of the spectrum by setting the bolometric correction in the 2--10 keV band to $f_{X} = 0.1$ \citepalias{Hasinger2020}.  

Given the specific luminosity, $L_{\nu}$, the specific emissivity of a population of PBHs accreting in the IGM is
\begin{equation}
    \dot{\rho}_{\text{IGM}}(\nu,z) = n_{\text{PBH}}^{\text{IGM}}(z)L_{\nu}(z).
    \label{eq:emissivity_IGM}
\end{equation}
The specific luminosity of an entire halo can be computed by:
\begin{equation}
    L_{\nu}^h(z) = \int_0^{R_{\text{vir}}}\dd r \dv{N_{\text{PBH}}}{r}L_{\nu}(r,z).
    \label{eq:L_bol_halo}
\end{equation}
The specific emissivity of a population of PBHs accreting inside haloes is then given by integrating over the halo mass function \citep{Murray+2013}:
\begin{equation}
    \dot{\rho}_{h}(\nu,z) = \int_{M_{\text{min}}}^{M_{\text{max}}}\dd M L_{\nu}^h(M)\dv{n}{M},
    \label{eq:emissivity_haloes}
\end{equation}
where $M_{\text{max}} = M_h(T_{\text{vir}} = 10^4)$ is the minimum mass of haloes inside which stars can form and  $M_{\text{min}}$ is the minimum mass of haloes required to form a baryon overdensity \citep{Barkana+2001}:
\begin{equation}
    M_{\text{min}}(T_{\text{IGM}},z) = 1.3 \times 10^3~\rm M_{\sun}\Big(\frac{10}{1+z}\Big)^{3/2}\Big( \frac{T_{\text{IGM}}}{1~\text{K}}\Big)^{3/2}.
    \label{eq:M_min}
\end{equation}
Inside haloes with $M_h < M_{\text{min}}$, the gas density is close to the mean IGM one and therefore we consider their contribution in the IGM emissivity. 

The background intensity in a given band [$\nu_1$,$\nu_2$] is related to the specific emissivity by \citep{Fernandez+2010,Yue+2013}:
\begin{equation}
    I^{[\nu_1,\nu_2]} = \frac{c}{4\pi}\int\dd z \frac{\int_{\nu_1}^{\nu_2}\dd \nu\varepsilon_{\nu'}(z)}{H(z)(1+z)},
    \label{eq:background}
\end{equation}
where $\nu' = (1+z)\nu$ and $H(z)$ is the Hubble parameter as a function of redshift. The angular power spectrum of NIRB fluctuations from PBHs can be decomposed in a two-halo and a shot-noise term:
\begin{equation}
    C_l = C_l^{\text{2-halo}} + C_l^{\text{SN}}.
\end{equation}
The clustering component at frequency $\nu$ and for the multiple moment $l$ is given by \citep{Cooray+2004,Fernandez+2010}
\begin{equation}
    C_l^{\text{2-halo}} = \frac{c}{4\pi}\int\dd z \frac{\epsilon^{2}_{\nu^{'}}(z)}{H(z)r^2(z)(1+z)^2}P\Big(k = \frac{l}{r(z)},z\Big),
    \label{eq:C_l}
\end{equation}
where $r(z)$ is the comoving distance and $P(k,z)$ is the power spectrum of the underlying matter distribution. PBHs in the IGM correspond to DM in the linear regime and therefore $P_{\text{IGM}}(k,z) = P_{\text{lin}}(k,z)$, where the right hand side is the linear matter power spectrum. Instead, haloes are biased tracers of the linear matter density field and their power spectrum can be written as  $P^{h}(k,z) = b_{\text{eff}}(z)P(k,z)$, where the effective bias $b_{\text{eff}}$ is given by:
\begin{equation}
    b_{\text{eff}}(z) = \int_{M_{\text{min}}}^{M_{\text{max}}}\dd M b_{h}(M,z)\dv{n}{M} / \int_{M_{\text{min}}}^{M_{\text{max}}}\dd M \dv{n}{M},
\end{equation}
where $b_h(M,z)$ is the halo bias, as derived in \citet{Tinker+2010}. 

The shot noise angular power spectrum is described by the following equation \citep{Cooray+2012,Yue+2013}:
\begin{equation}
    C_l^{\text{SN}} = \frac{c}{(4\pi)^2}\int  \frac{\dd z}{H(z)r^2(z)(1+z)^2}\int_{M_{\text{min}}}^{M_{\max}} \dd M L_{\nu}^2(M)\dv{n}{M}.
    \label{eq:shot_noise}
\end{equation}
We note that in principle one should consider the one-halo term, given by \citep{Cooray+2012}:
\begin{align}
    C_l^{\text{1-halo}} = \frac{c}{(4\pi)^2} &\int  \frac{\dd z}{H(z)r^2(z)(1+z)^2} \\
    &\int_{M_{\text{min}}}^{M_{\max}} \dd M L_{\nu}^2(M)\dv{n}{M}\abs{\Tilde{u}(k = l/r(z),M)}^2,
    \label{eq:one_halo}
\end{align}
where $\Tilde{u}(k = l/r(z),M)$ is the Fourier transform of the NFW profile. For the redshift and halo mass range of interest, we checked that $\Tilde{u}(k = l/r(z),M) \sim 1$ and therefore the one-halo term reduces to the shot noise term in equation~(\ref{eq:shot_noise}).

\subsection{Matter Power spectrum modified by PBHs}
\label{subsec:PBH-LCDM}
 PBHs may constitute a fraction of DM, thus they would add a Poisson shot noise term to the linear matter power spectrum \citep{Meszaros1975,Afshordi+2003}:
 \begin{equation}
     P_{\text{poiss}} = \frac{f_{\text{PBH}}^2}{n_{\text{PBH,0}}},
 \end{equation}
 where $n_{\text{PBH,0}}$ is the PBH number density at redshift $z = 0$. The total matter power spectrum can be then written as \citep{Villanueva-Domingo+2023}:
\begin{equation}
    P_{\text{PBH}-\Lambda \text{CDM}}(z,k) = P_{\Lambda \text{CDM}}(k,z) + D^2(z)T^2_{\text{iso}}(k)P_{\text{poiss}},
    \label{eq:P_tot}
\end{equation}
where $D(z)$ is the linear growth factor and $T_{\text{iso}}$ is the isocurvature transfer function. An approximate expression for $T_{\text{iso}}$ is given by \citep{Peacock1998}: 
\begin{equation}
T_{\text{iso}} = 
    \begin{cases}
    \frac{3}{2}(1+z_{\text{eq}}), & k \geq k_{\text{eq}} \\
    0, &k < k_{\text{eq}}
    \end{cases}
\end{equation}
 where $z_{\text{eq}}$ is the redshift of radiation-matter equality and $k_{\text{eq}} = c^{-1}H(z_{\text{eq}})/(1+z_{\text{eq}})$.
 The contribution to the power spectrum from PBHs can be recast in the form \citep{Villanueva-Domingo+2023}:
\begin{equation}
    P_{\text{PBH}} = T^2_{\text{iso}}P_{\text{poiss}} = 2.5\times 10^{-2}\,f_{\text{PBH}} \Big(\frac{M_{\text{PBH}}}{30~\rm M_{\sun}} \Big)~\text{Mpc}^3.
    \label{eq:P_PBH}
\end{equation}
The PBH modification to the power spectrum affects the variance of the matter overdensity field and thus the halo mass function. 

\begin{figure}
    \centering
    \includegraphics[width = \columnwidth]{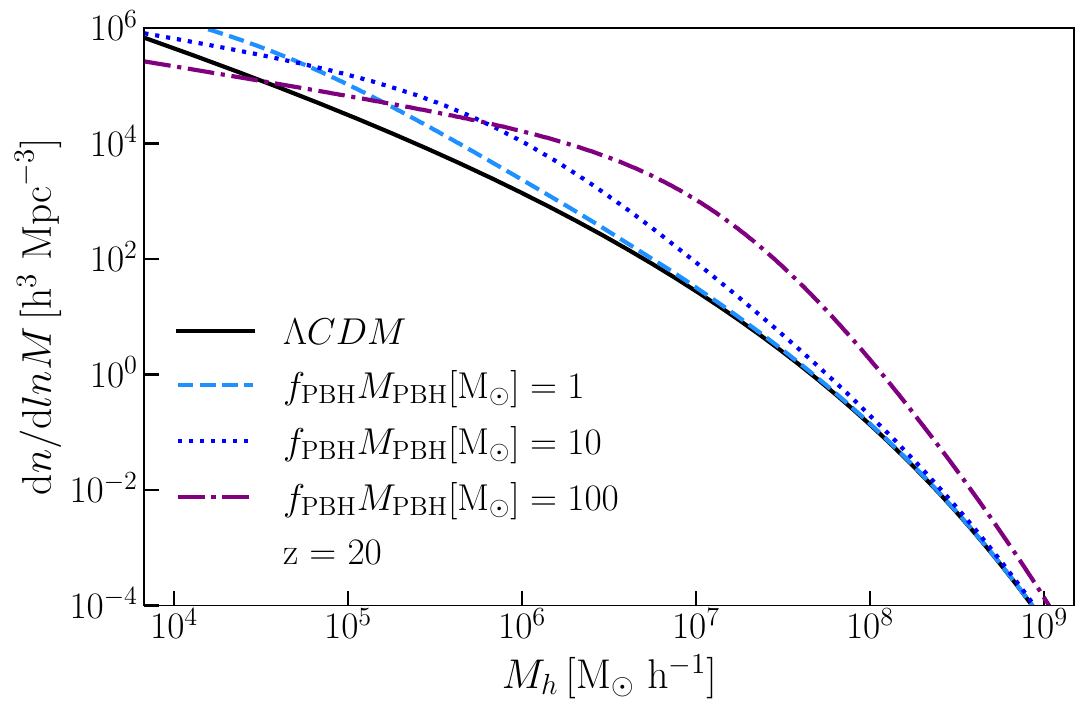}
    \caption{Press--Schechter halo mass function at $z = 20$, in the standard $\Lambda$CDM scenario (black solid line) and including the modification induced by PBHs, with $f_{\text{PBH}}M_{\text{PBH}} = 1~\rm M_{\sun}$ (light blue dashed),  $f_{\text{PBH}}M_{\text{PBH}} = 10~\rm M_{\sun}$ (blue dotted) and $f_{\text{PBH}}M_{PBH} = 100~\rm M_{\sun}$ (purple dot-dashed).}
    \label{fig:hmf_PBH}
\end{figure}

\begin{figure}
    \centering
    \includegraphics[width = \columnwidth]{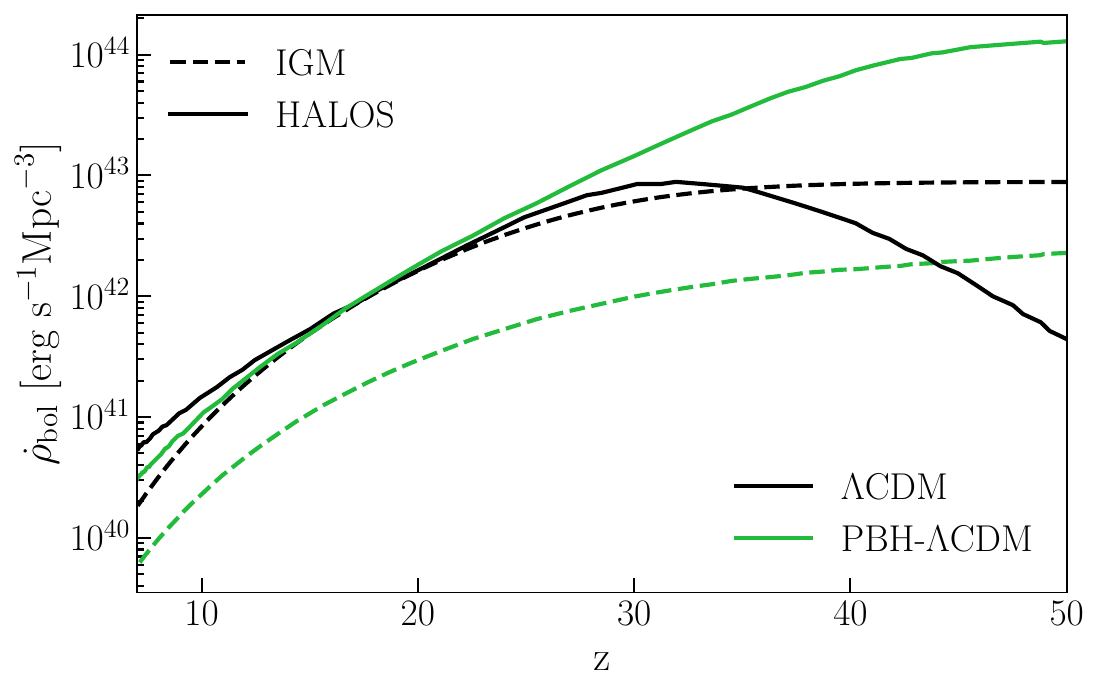}
    \caption{Bolometric emissivity from IGM (dashed lines) and haloes (solid lines), in the standard $\Lambda$CDM scenario (black) and including the power spectrum modified by PBHs (green). In the case of PBH-$\Lambda$CDM cosmology, the halo signal dominates over the IGM one for all the redshift of interest. Here we adopt $M_{\text{PBH}} = 30~\rm M_{\sun}$.}
    \label{fig:em_bol_PBH}
\end{figure}

{Hereafter we will refer to a PBH-$\Lambda$CDM cosmology whenever adopting the power spectrum described by equations~(\ref{eq:P_tot})--(\ref{eq:P_PBH}). In particular we consider a PBH-$\Lambda$CDM cosmology in our models \modB{} and \modC{} (see Sec.~\ref{sec:Results}).} In Fig.~\ref{fig:hmf_PBH} we show the Press--Schechter halo mass function at $z = 20$ in the standard $\Lambda$CDM scenario and including the modification induced by PBHs, for different values of the parameter $f_{\text{PBH}}M_{\text{PBH}}$. At $z = 20$, when including the extra power on small scales due to PBHs, the halo mass function is a factor of 3 (40) higher for $M_h = 10^{5}~\rm M_{\sun}$ ($M_h = 10^{7}~\rm M_{\sun}$) with respect to the standard $\Lambda$CDM case, considering $f_{\text{PBH}}M_{\text{PBH}} = 100$ .
In Fig.~\ref{fig:em_bol_PBH} we compare the bolometric emissivity, from both haloes and IGM, in the $\Lambda$CDM and PBH-$\Lambda$CDM cosmologies, as a function of redshift. As a consequence of the increased number of small haloes expected in the PBH-$\Lambda$CDM, the contribution to the total emissivity from accreting PBHs in DM haloes is enhanced by a factor of 2 (20) at redshift $z = 30$ (40). Moreover, also the collapsed DM fraction is higher and thus the relative contribution from PBHs accreting in the IGM is further lowered. The emissivity of PBHs accreting in haloes at redshift $z = 20~ (40)$ is roughly 10 (100) times the emissivity from PBHs in the IGM. We point out that, as a consequence of the aforementioned effects, in the PBH-$\Lambda$CDM cosmology halo emissivity dominates the IGM one at any redshift, unlike in the standard $\Lambda$CDM case.

\subsection{Extended PBH mass function}
\label{subsec:ext_mass_func}
The mass function of PBHs at the epoch of their formation is denoted by $\psi(M_{\text{PBH}})$, and defined as:
\begin{equation}
    \psi(M_{\text{PBH}}) = \frac{1}{f_{\text{PBH}}}\dv{f_{\text{PBH}}}{M_{\text{PBH}}}.
\end{equation}
In the following, we generalize our formalism to extended mass functions. In particular, we consider the case of a lognormal mass function\footnote{We choose a lognormal mass function to avoid fruitless complications. The main results of our work are unaffected by the exact shape of the mass function.}:
\begin{equation}
    \psi_{\text{log}}(M) = \frac{1}{M\sqrt{2\pi}\sigma}\exp\Big(-\frac{\log(M/M_c)^2}{2\sigma^2}\Big).
\end{equation}
Here $M_c$ is the critical mass that sets the position of the peak and $\sigma$ is the standard deviation of the distribution. 
Regarding PBHs in the IGM, their emissivity can be generalized to
\begin{equation}
    \dot{\rho}_{\text{IGM}}(\nu,z) = \int \dd M n_{\text{PBH}}^{\text{IGM}}(z,M) \psi(M)L_{\nu}(z,M),
\end{equation}
where the integral is performed over the PBH mass and $n_{\text{PBH}}^{\text{IGM}}(z,M)$ is taken from equation~(\ref{eq:n_PBH_IGM_HALO}). Recalling that $n_{\text{PBH}}^{\text{IGM}} \propto M^{-1}$ and $L_{\nu} \propto M^2$, we can write
\begin{equation}
   \dot{\rho}_{\text{IGM}} \propto \int \dd M \psi(M)M \equiv \bar{M}.
\end{equation}
Therefore, when computing the emissivity of PBHs in the IGM, an extended mass function is equivalent to a $\delta$-function centered at the mean mass $\bar{M}$ of the mass function\footnote{This is valid for a constant radiation efficiency. If $\varepsilon \propto \dot{M}^a \propto M^{2a}$, then $L_{\nu} \propto M^{2+2a}$ and so the corresponding mean mass should be $\bar{M} = \Big( \int \dd M\psi(M)M^{2a+1}\Big)^{1/(2a+1)}$.}. For a lognormal mass function, the mean mass is $\bar{M}_{\text{log}} = M_c\exp(\sigma^2/2)$. We will assume the benchmark value $\sigma = 1$ throughout the rest of the paper and quote only the mean mass of the lognormal distribution.

Regarding PBHs accreting in haloes, we must specify how PBHs of different masses are distributed inside the halo. {We note that the frictional acceleration exerted onto a body of mass $M$ moving through a homogeneous distribution of particles of mass $m$ ($m \ll M$) with isotropic velocity distribution is $\propto M$ \citep{Binney&Tremaine2008}. Hence, PBHs with higher masses sink towards the centre of the halo before lighter ones.}
With this in mind and for simplicity, we then assume that more massive PBHs lie at smaller radii.  

\begin{figure}
    \centering
    \includegraphics[width = \columnwidth]{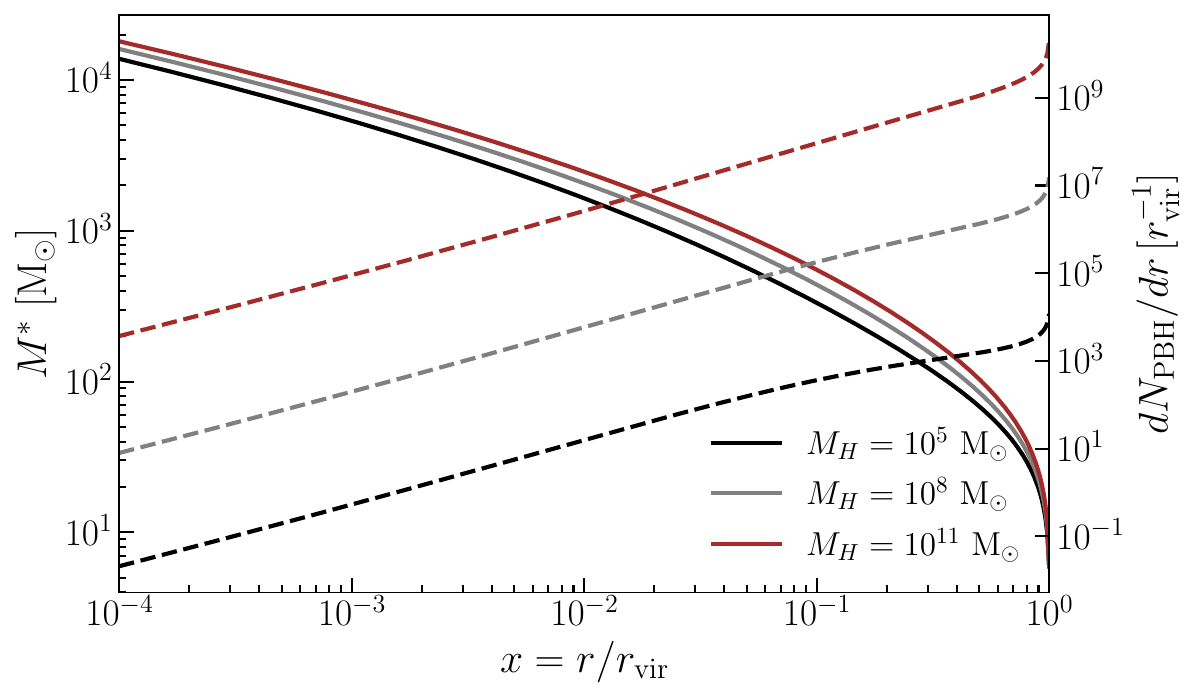}
    \caption{Mass of PBHs present at radius $r$ in haloes of masses $10^{5}\,\rm M_{\sun}$ (black), $10^{8}\,\rm M_{\sun}$ (grey), $10^{11}\,\rm M_{\sun}$ (brown), when considering a lognormal mass function with $\bar{M}_{\text{log}} = 30~\rm M_{\sun}$. The dashed lines show the number of PBHs per unit length, in units of $r_{\text{vir}}^{-1}$.}
    \label{fig:M_PBH_star}
\end{figure}

\begin{figure}
    \centering
    \includegraphics[width = \columnwidth]{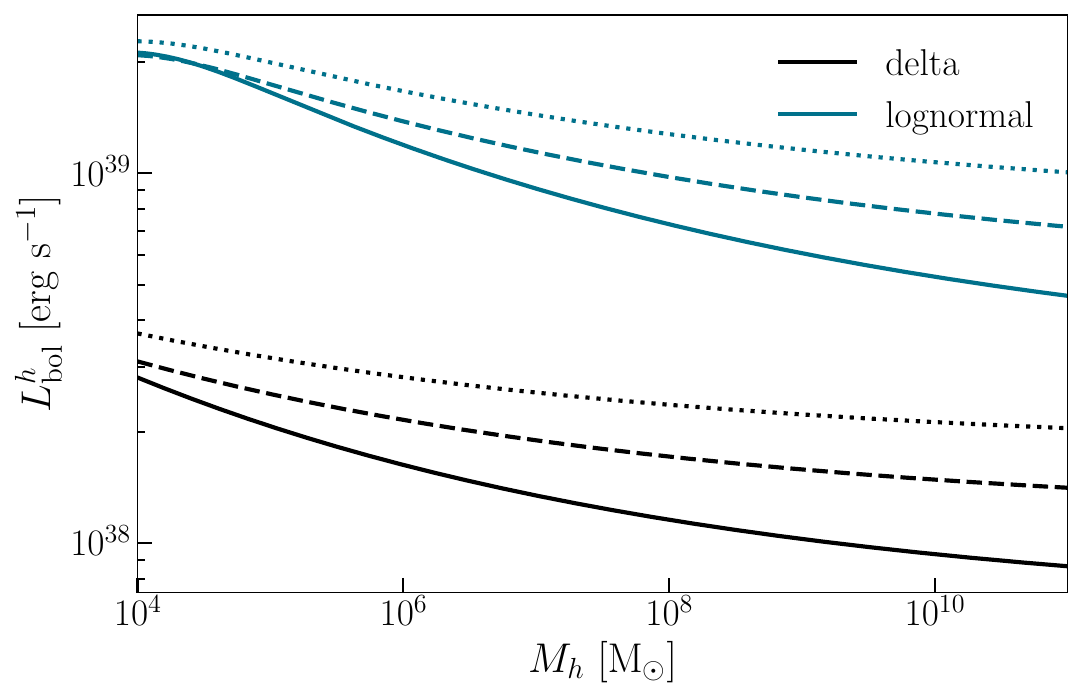}
    \caption{Bolometric luminosity of haloes as a function of the halo mass, in the case of a delta (black) and lognormal (blue) mass function, for $z = 20,30,40 $ (solid, dashed, dotted respectively). The PBH mass for the delta mass function and the mean mass of the lognormal distribution are both set to $M_{\text{PBH}} = \bar{M}_{\text{log}} = 30~M_{\odot}$. With an extended mass function, the luminosity is boosted by a factor $\sim 10$, because more massive PBH tend to sink towars the center where the gas density is higher. }
    \label{fig:Lbol_Mh_log}
\end{figure}

The mass $M^{*}(r)$ of PBHs at a given radius $r$ can be derived by imposing that the mass $M(r)$ enclosed in a sphere of radius $r$ is equal to the integrated mass of all PBHs more massive than $M^{*}(r)$:
\begin{equation}
   M(r) = 4\pi\int^{r}_0 \dd r' {r'}^2 \rho_{\text{NFW}}(r') = \int_{M^{*}(r)}^{\infty} \dd M' M' \psi(M').
    \label{eq:M_PBH_star}
\end{equation}
For a lognormal mass function, the right-hand side of the above equation can be computed analitically, giving
\begin{equation}
    \log[M^{*}(r)/M_c] = \sigma^2+\sqrt{2}\sigma\erf^{-1}\Big[1-\frac{2M(r)}{M_h}\Big],
\end{equation}
where $M_h$ is the halo mass. We show the resulting $M^{*}(r)$ for a lognormal mass function with $\bar{M}_{\text{log}} = 30~\rm M_{\sun}$ in Fig.~\ref{fig:M_PBH_star}.
Once specified $M^{*}(r)$, the number of PBHs per unit length at radius $r$ is given by
\begin{equation}
    \dv{N_{PBH}}{r} = 4\pi \frac{f_{PBH}}{M^{*}(r)}\rho_{DM}(r).
    \label{eq:N_PBH_ext_mass_func}
\end{equation}
We can then substitute equation~(\ref{eq:N_PBH_ext_mass_func}) into equation~(\ref{eq:L_bol_halo}) and apply the same formalism described in Sec.~\ref{subsec:NIRB}. In Fig.~\ref{fig:Lbol_Mh_log} we compare the bolometric luminosity of haloes in the case of a delta mass function with $M_{\text{PBH}} = 30~\rm M_{\sun}$ and of a lognormal mass function with $\bar{M}_{\text{log}} = 30~M_{\odot}$. Including the lognormal mass function boosts the halo luminosity by a factor of $\sim 10$, because more massive PBHs accrete at smaller distances from the center, where the gas density is higher.

\section{Results}
\label{sec:Results}
In this section we present the IGM temperature evolution, the mean NIRB intensity and the NIRB angular power spectrum obtained from three different models: (\textit{i}) standard $\Lambda$CDM cosmology with a PBH delta mass function ($\Lambda$CDM-$\delta$); (\textit{ii}) PBH-$\Lambda$CDM cosmology with PBH delta mass function (PBH-$\delta$) and (\textit{iii}) PBH-$\Lambda$CDM cosmology with a PBH lognormal mass function (PBH-lognormal). 
We compare our predictions to observational data to test the hypothesis of accreting PBHs as sources of the NIRB. 

\subsection{IGM temperature and ionization evolution}
\label{subsec:IGM_temperature_evolution}
X-ray emission from PBHs would heat and ionize the IGM well before galaxies start to reionize the Universe. To account for this effect, we self-consistently derive the IGM temperature and ionization evolution following the formalism described in Sec. 3 of \citetalias{Ziparo+2022}. 

In the pre-overlap phase of the cosmic reionization process, the Universe can be split in ionized and neutral regions. In the ionized regions, the redshift evolution of the free electron fraction $x_e(z)$ is solved through equation (33a) in \citetalias{Ziparo+2022}, adopting the photoionization rate derived from the UV background in \citet{Puchwein+2019}. The free electron fraction traces the evolution of the volume filling factor of ionised regions, namely the fraction of volume occupied by ionized regions. In the same regions, we assume an IGM temperature $T_{\text{IGM,ion}} = 10^4$ K. Such high temperature suppresses accretion onto PBHs due to high sound speeds (equation~ (\ref{eq:sound_speed_IGM})). 

In neutral regions, whose volume filling factor is $1-x_e(z)$, the free electron fraction $x_{e,n}(z)$ evolves with redshift according to equation (35a) in \citetalias{Ziparo+2022}. 
Here, the photoionization rate calculation accounts for secondary ionizations due to X-rays emitted by PBHs:
\begin{equation}
    \Gamma_{\text{PBH}} = \int_{\nu_{\text{min}}}^{\infty}\dd\nu\frac{4\pi I_{\nu}}{h\nu}\left(\frac{h\nu}{E^{\text{th}}}-1\right)f_{\text{ion}}\sigma_H(\nu),
\end{equation}
where $h\nu_{\text{min}} = 0.5~\text{keV}$ \footnote{We neglect the contribution of UV photons to IGM heating and ionization since, as discussed in \citetalias[Appendix B]{Ziparo+2022}, the Stromgren sphere surrounding accreting PBHs results to be comparable to the Bondi radius, preventing UV photons to contribute to the IGM ionization.}, $E^{\text{th}}$ is the hydrogen ionization threshold, $\sigma_H$ is the hydrogen ionization cross section and $f_{\text{ion}} \sim 0.3$  is the fraction of the primary electron's energy going into secondary ionizations \citep{Furlanetto+2010}. In neutral regions we also solve the redshift evolution of temperature $T_{\text{IGM,n}}(z)$ through equation (35b) in \citetalias{Ziparo+2022}, which takes into account the IGM heating due to the energy injected by X-rays \citep{Mesinger+2013}, here assumed to be emitted by PBHs. The heating rate $\epsilon_{\text{PBH}}$ per baryon can then be computed as:
\begin{equation}
    \epsilon_{\text{PBH}} = \int_{\nu_{\text{min}}}^{\infty}\dd\nu\frac{4\pi I_{\nu}}{h\nu}(h\nu-E^{\text{th}})f_{\text{heat}}\sigma_H(\nu),
\end{equation}
where $f_{\text{heat}} \sim 0.3$ is the fraction of the primary electron's energy going into heat \citep{Valdes10}.

Following the evolution of neutral regions $T_{\text{IGM,n}}$ is crucial because changes in the IGM temperature affects PBHs emissivity. On the one hand, if the IGM temperature increases, the effective velocity of PBHs accreting in the IGM increases as well (equation~(\ref{eq:v_eff})): this lowers their luminosity and consequently their emissivity. On the other hand, to higher $T_{\text{IGM,n}}$ correspond higher $M_{\text{min}}$ (equation~(\ref{eq:M_min})): the integration interval in equation~(\ref{eq:emissivity_haloes}) is thus shortened, which reduces the emissivity of PBHs accreting in haloes. Therefore, if $T_{\text{IGM,n}}$ increases (decreases), the total emissivity of PBHs is lowered (enhanced). 

\begin{figure*}
    \centering
    \includegraphics[width = \textwidth]{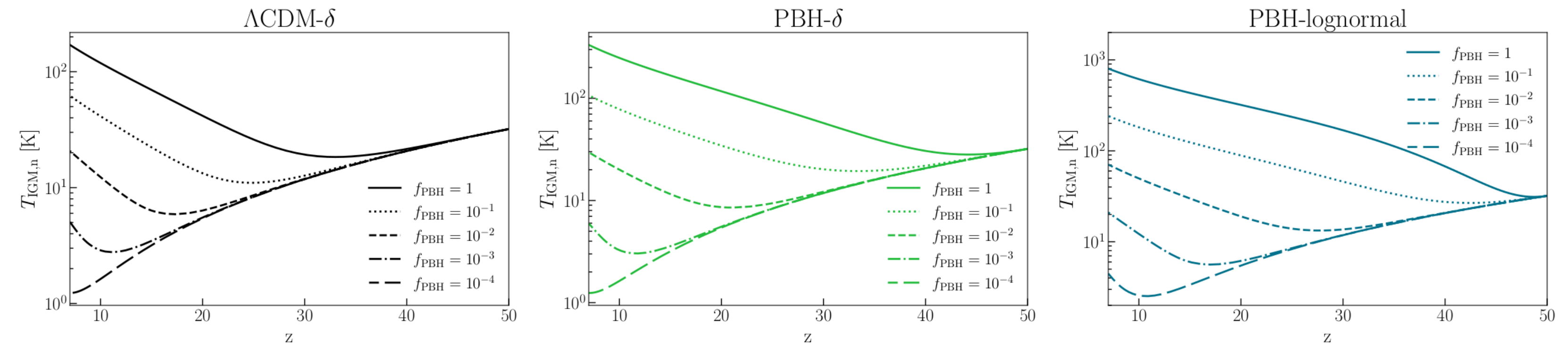}
    \caption{Evolution of the neutral IGM temperature as a function of redshift in the three different models \modA{}, \modB{} and \modC{} (black, green and blue respectively). Each line corresponds to different values of $f_{\text{PBH}}$: $10^{-4}$, $10^{-3}$, $10^{-2}$, $10^{-1}$, $1$ (long-dashed, dot-dashed, dashed, dotted and solid respectively). The PBHs mass is  $M_{\text{PBH}} = \bar{M}_{\text{log}} = 30~\rm M_{\sun}$. We maintain the same color-coding for the models throughout the rest of the paper.}
    \label{fig:T_IGM}
\end{figure*}

We show the resulting temperature evolution in Fig.~\ref{fig:T_IGM} for the three different models, adopting $M_{\text{PBH}} = \bar{M}_{\text{log}} = 30~\rm M_{\sun}$ and $f_{\text{PBH}} = 10^{-4},10^{-3},10^{-2},10^{-1},1$. 
For $f_{\text{PBH}} = 1$, heating from PBHs increases the IGM temperature in neutral regions at $z\sim 6$ by a factor of $\sim 10, 20, 60$ in model $\Lambda$CDM-$\delta$, PBH-$\delta$, PBH-lognormal respectively, with respect to a $\Lambda$CDM cosmology which does not include PBHs. While in model $\Lambda$CDM-$\delta$ $T_{\text{IGM,n}}$ starts increasing around $z \sim 30$, in model PBH-$\delta$ it rises at higher redshifts ($z \sim 40$), as the PBH emissivity is boosted by the higher number of small mass ($M_h \lesssim 10^{6}$--$10^{7}~\rm M_{\sun}$) haloes. For $f_{\text{PBH}} \lesssim 10^{-2}$,  the effect of PBHs on the halo mass function is negligible and the evolution of $T_{\text{IGM,n}}$ in the two cases is almost identical. In model PBH-lognormal, the luminosity of haloes is further enhanced by the lognormal mass function (Fig.~\ref{fig:Lbol_Mh_log}) and  $T_{\text{IGM,n}}$ reaches $\sim 900$K at $z\sim 6$.

Before moving to the core results of this work, we briefly comment on the implications of IGM heating from PBHs. Firstly, our model does not affect the IGM temperature at $z \lesssim 5$, where measurements from Lyman-alpha forest observations are obtained \citep{Walther+2019,Gaikwad+2020}. Below $ z\sim 6$, most of the Universe is ionized and IGM temperatures $T_{\text{IGM,ion}} \gtrsim 10^4$ suppress the emission and thus the heating from PBHs in ionized regions. Moreover, the contribution from PBHs in neutral regions is also suppressed because their volume filling factor, i.e $1-x_e$ in our model, approaches zero as cosmic reionization proceeds. Instead, at $z \gtrsim 10$, the Universe is mostly neutral and the volume filling factor of neutral regions is basically unity. Radiation from PBHs is then effective in heating the IGM above the adiabatic cooling temperature. Therefore, forthcoming 21 cm observations could provide stringent constraints on the heating from PBHs and hence on their abundance \citep{Mena+2019}.

\subsection{NIRB mean intensity and angular power spectrum}
As already mentioned in the Introduction, direct measurements of the mean NIRB intensity are uncertain \citep{LibraeReview}. However, constraints on the mean NIRB excess from unknown sources can be derived from the angular power spectrum measurement, as done by \citet{Kashlinsky+2007}, who found $I^{2-5\,\micron} \gtrsim 1~\text{nW m}^{-2} \text{sr}^{-1}$ in the 2--5 $\micron$ band.
\label{subsec:NIRB_results}
\subsubsection{NIRB mean intensity}
\begin{figure}
    \centering
    \includegraphics[width = \columnwidth]{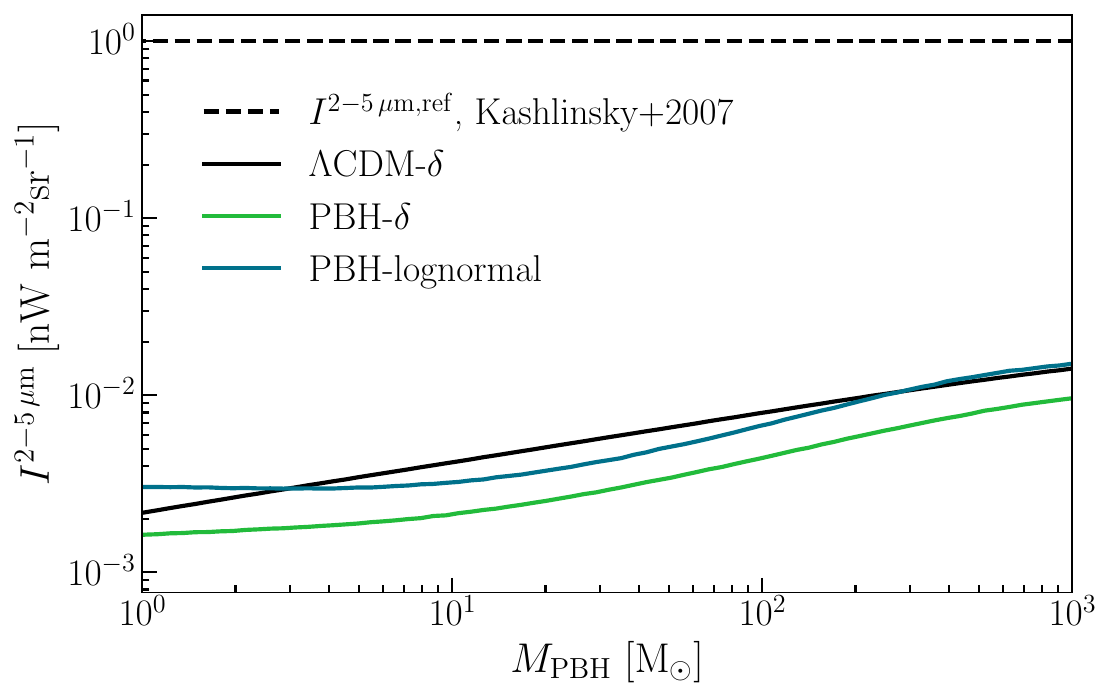}
    \caption{Near infrared background intensity in the band 2-5 $\micron$ predicted by the three different models, as a function of PBH mass. The horizontal dashed line corresponds to the minimal near-infrared flux required by NIRB fluctuations \citep{Kashlinsky+2007}.}
    \label{fig:NIR_flux}
\end{figure}

We compute the mean NIRB intensity in the 2--5 $\micron$ band, using equation~(\ref{eq:background}). To test our predictions, in Fig. \ref{fig:NIR_flux} we compare these results with data from \citet{Kashlinsky+2007}. We consider the limiting case $f_{\text{PBH}}  = 1$, which sets the upper limit for NIR flux produced by PBHs, for a mass range $1~\rm M_{\sun} \le M_{\text{PBH}} = \bar{M}_{\text{log}} \le 10^{3}~\rm M_{\sun}$. Our choice is driven, one the one hand, by the requirement of substantial accretion rates and hence luminosities and, on the other hand, by existing constraints on the abundance of higher mass PBHs. We show the results for the three different models in Fig.~\ref{fig:NIR_flux}. We find that in the $\Lambda$CDM-$\delta$ (PBH-$\delta$, PBH-lognormal) model, PBHs contribution to the NIRB mean intensity is at most 1.4 (0.9,1.5) per cent, if $M_{\text{PBH}} = 10^3~\rm M_{\sun}$.

\begin{figure}
    \centering
    \includegraphics[width = \columnwidth]{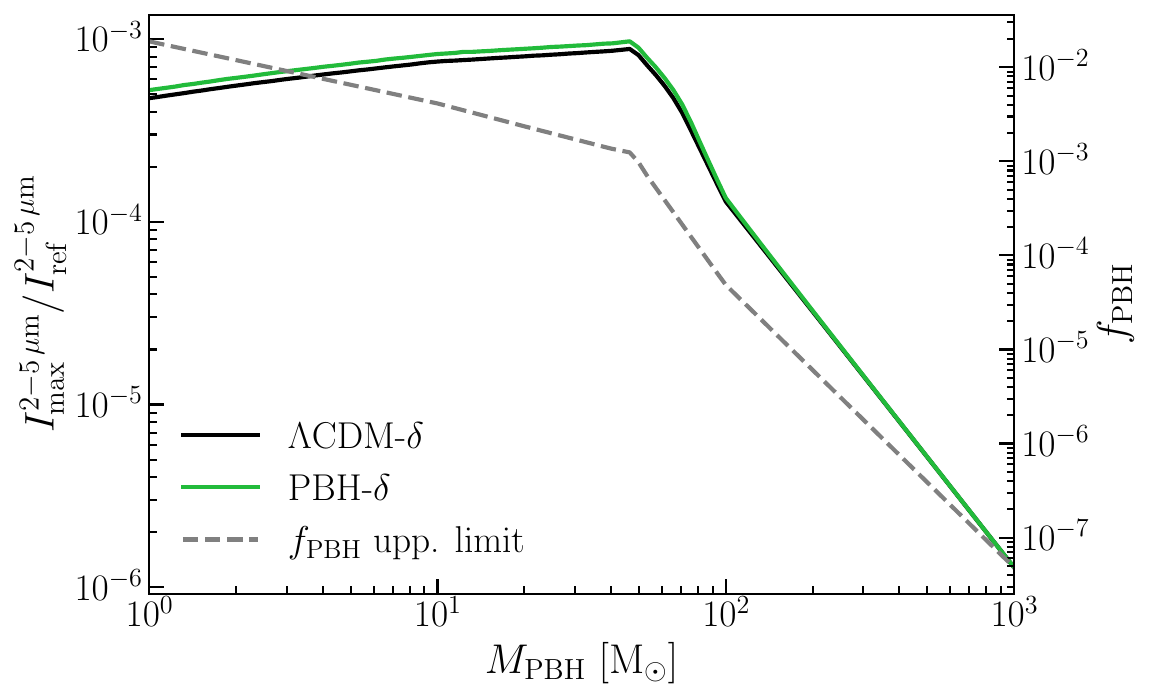}
    \caption{Maximum intensity of the NIRB produced by PBHs accounting for present constraints, in units of the reference value ($1~\text{nW m}^{-2} \text{sr}^{-1}$), for model $\Lambda$CDM-$\delta$ (solid black) and \modB{} (solid green). We do not show the \modC{} model, because available upper limits on $f_{\text{PBH}}$ are derived only for a $\delta$-mass function. We also plot the strongest constraint on $f_{\text{PBH}}$ (dashed grey) as a function of the PBH mass. We use the LIGO and CMB bounds from PBHBounds \citep{PBH_bounds}.}
    \label{fig:flux_NIR_upper_limit}
\end{figure}

However, current constraints on the abundance of PBHs \citep{Carr+2020} already exclude $f_{\text{PBH}} = 1$ in the mass range considered here. Lower values of $f_{\text{PBH}}$ result in lower NIR flux produced by PBHs. In Fig.~\ref{fig:flux_NIR_upper_limit} we show, as a function of PBH mass, the ratio between the intensity of the NIRB produced by PBHs computed considering the maximum value of $f_{\text{PBH}}$ allowed by existing constraints and the NIRB mean intensity reference value. We also show the most stringent upper limit on $f_{\text{PBH}}$, which in the mass range of interest are derived from LIGO observations of black holes mergers \citep{LIGO2017,Kavanagh+2018} and from CMB angular power spectrum \citep{Poulin+2017}.

We find that, in the most favorable case consistent with constraints, PBHs can produce 0.1 per cent of the NIRB intensity if $M_{\text{PBH}} \sim 50~\rm M_{\sun}$.  
We show the results only for the \modA{}  and \modB{} models, as we consider constraints computed adopting a delta mass function. We note that the difference between the two models is very tiny, as for low values of $f_{\text{PBH}}$ the modification induced by PBHs to the matter power spectrum is almost negligible.

\subsubsection{NIRB angular power spectrum}
\begin{figure}
    \centering
    \includegraphics[width = \columnwidth]{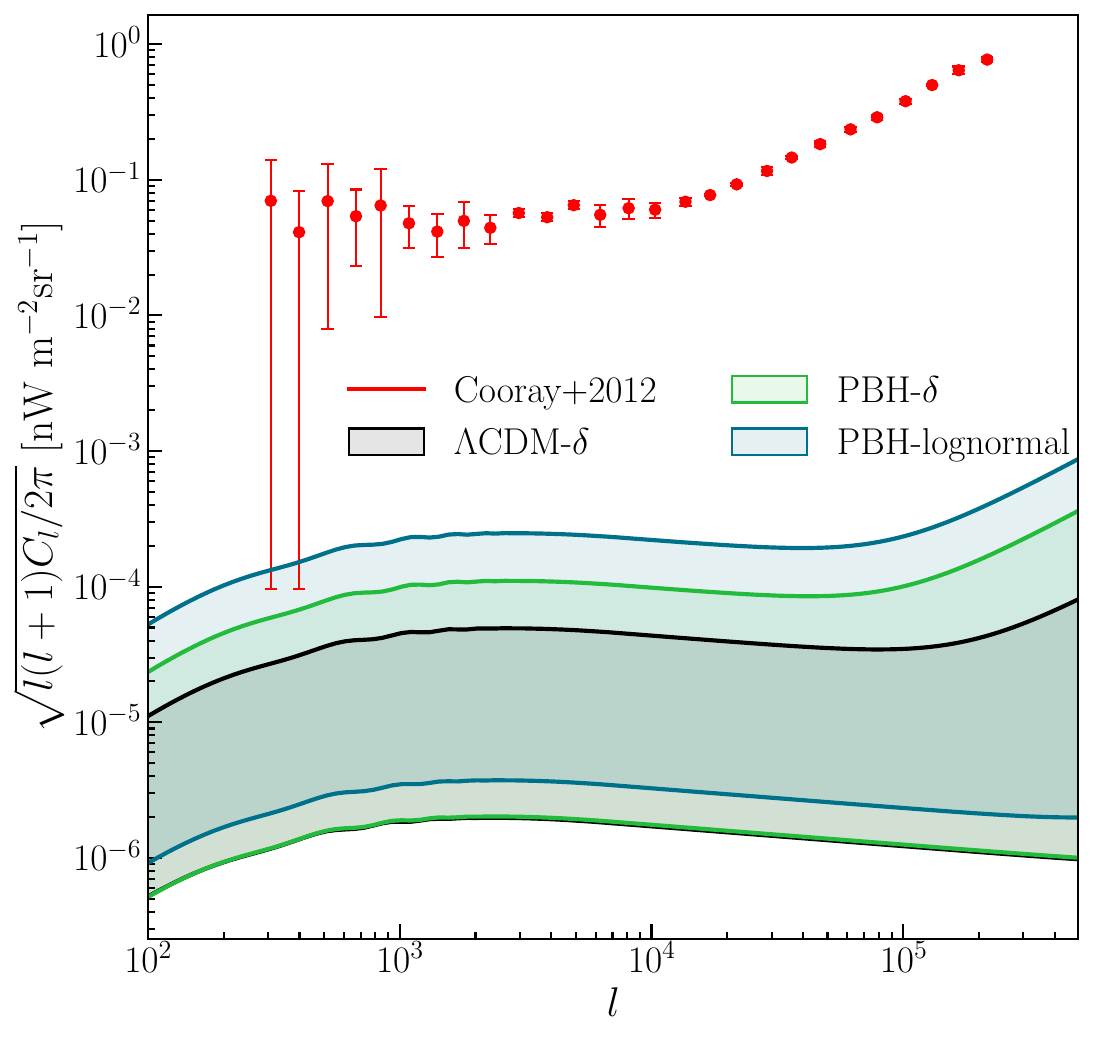}
    \caption{Angular power spectrum of the NIRB at 3.6 $\mu$m from PBHs for the three models, compared with observational data from \citet{Cooray+2012}. Shaded regions correspond to values of PBH mass between $1~\rm M_{\sun}$ (lower lines) and $ 10^{3}~\rm M_{\sun}$ (upper lines).  }
    \label{fig:Cl_3.6}
\end{figure}

We also compute the angular power spectrum of NIRB fluctuations produced by PBHs at the reference wavelength 3.6 $\micron$, using equations.~(\ref{eq:C_l}) and (\ref{eq:shot_noise}). We show the results in Fig. \ref{fig:Cl_3.6}, for the case $f_{\text{PBH}} = 1$ and with PBH masses between $1~\rm M_{\sun}$ (lower lines) and $ 10^{3}~\rm M_{\sun}$ (upper lines). We compare our predictions with the latest measurements of NIRB angular power spectrum from Spitzer Deep, Wide-Field Survey \citep{Cooray+2012}. 

None of the models considered in this work is able to reproduce the observed angular power spectrum. At multiple moment $l \sim 10^3$, corresponding to angular scales of $\theta = 2\pi/l \sim 20$ arcmin, fluctuations predicted by the \modA{} (\modB{}, \modC{}) model are lower than the measured one by a factor of $1000~(400, 200)$. This holds for $M_{\text{PBH}} = 10^3~\rm M_{\sun}$, which provides the highest value of the angular power spectrum in the mass range considered here.

\section{Summary and Discussion}
\label{sec:Conclusion}
In this work, we have tested the hypothesis that PBHs are sources of the NIRB excess. By assuming that PBHs constitute a fraction $f_{\rm PBH}$ of cold dark matter (DM), we have computed the mean intensity and angular power spectrum of the NIRB arising from their accretion. 

Following the formalism by \citet{Ziparo+2022}, we account for PBH accretion both in the intergalactic medium (IGM) and in DM haloes, and we self-consistently derive the IGM temperature evolution, considering ionization and heating due to X-ray emission from PBHs. The \citetalias{Ziparo+2022} model is based on the $\Lambda$CDM linear matter power spectrum, and considers a $\delta$ function for the PBH mass distribution. 

Besides this $\Lambda$CDM-$\delta$ model, we have considered the possibility that PBHs modify the matter power spectrum (PBH-$\delta$ model), and follow an extended lognormal mass function (PBH-lognormal model). {In both \modB{} and \modC{} models we adopt a PBH-$\Lambda$CDM cosmology, accounting for the matter power spectrum modified by PBHs.} 

For each model, we have derived the intensity and angular power spectrum of the NIRB finding that PBHs contribute to the observed NIRB fluctuations to $<1$ per cent, even in the most optimistic cases considered in this work. This conclusion is supported by these intermediate results:
\begin{itemize}
\item The PBH modification to the power spectrum affects the variance of the matter overdensity field and thus the halo mass function, adding an extra power on small scales. In particular, at $z = 20$, in the PBH-$\Lambda$CDM model, the halo mass function is a factor of 3 (40) higher for $M_h = 10^{5}~\rm M_{\sun}$ ($M_h = 10^{7}~\rm M_{\sun}$) with respect to the standard $\Lambda$CDM case, considering $f_{\text{PBH}}M_{\text{PBH}} = 100$ .

\item As a consequence of the increased number of small haloes expected in the PBH-$\Lambda$CDM, the contribution to the total emissivity from accreting PBHs in DM haloes is enhanced by a factor of 2 (20) at redshift $z = 30$ (40). Moreover, also the collapsed DM fraction is higher and thus the relative contribution from PBHs accreting in the IGM is further lowered. The emissivity of PBHs accreting in haloes at redshift $z = 20~ (40)$ is roughly 10 (100) times the emissivity from PBHs in the IGM. We point out that, as a consequence of the aforementioned effects, in the PBH-$\Lambda$CDM cosmology the halo emissivity dominates the IGM one at any redshift, unlike in the standard $\Lambda$CDM case.

\item If the radiative efficiency is indipendent of the accretion rate, given an extended mass function, the emissivity of PBHs accreting in the IGM can be computed adopting a delta mass function with mass equal to the mean mass of the mass function $\bar{M}_{\text{log}}$. We compare the bolometric luminosity of haloes in the case of a delta mass function with $M_{\text{PBH}} = 30~\rm M_{\sun}$ and of a lognormal mass function with $\bar{M}_{\text{log}} = 30~M_{\odot}$. Including the lognormal mass function boosts the halo luminosity by a factor of $\sim 10$, because more massive PBHs accrete at smaller distances from the center, where the gas density is higher.

\item Considering $1 \le M_{\text{PBH}}~[\rm M_{\sun}] \le 10^3$ and $f_{\text{PBH}} = 1$, PBHs can produce at most $\sim$ 1 per cent of the flux required to explain NIRB fluctuations. The three models differ in their prediction by less than a factor of $\simeq$ 2. Although in PBH-$\Lambda$CDM cosmology the total emissivity of PBHs at $z\gtrsim 40$ is $\sim 10$ higher than in the standard $\Lambda$CDM scenario (for both delta and lognormal mass functions), the resulting NIRB is similar, because the gas heating from X-rays produced by PBHs damps their emissivity at lower redshifts. 

\item When accounting for current constraints on PBH abundance, the maximum relative contribution of PBHs to the NIRB is reduced to 0.1 per cent, for PBHs with $M_{\text{PBH}} \sim 50~\rm M_{\sun}$.

\item None of our models is able to reproduce the NIRB angular power spectrum. At large angular scales ($\theta \sim 20$ arcmin), fluctuations predicted by model $\Lambda$CDM-$\delta$ (PBH-$\delta$, PBH-lognormal) are lower than the measured one by a factor of $1000~(400, 200)$, in the most favorable case with $M_{\text{PBH}} = 10^3~\rm M_{\sun}$.         
\end{itemize}

Before concluding, we compare our findings with the results from \citet[hereafter \citetalias{Hasinger2020}]{Hasinger2020}, whose model is adopted in \citet{Cappelluti+2022}. \citetalias{Hasinger2020} predicted a NIR flux from PBHs of $10^{-13}~\text{erg s}^{-1}\text{cm}^{-2}\text{deg}^{-2} \sim 3 \times 10^{-4}~\text{nW m}^{-2} \text{sr}^{-1}$. This corresponds to $\sim 0.3$ per thousand of the NIRB flux required to explain NIRB fluctuations. Hence, we find a NIRB flux $\sim 10 \times$ higher then the one obtained in \citetalias{Hasinger2020}. We point out some substantial differences between the two models to fully grasp the discrepancy in the two results. Firstly, \citetalias{Hasinger2020} adopt a non-linear relative velocity between gas and DM to capture the collapse of baryons into DM haloes. This approach does not account for the density profile of DM and gas inside haloes, which enhance the contribution of PBHs accreting in haloes, as gas densities are much higher than the mean baryon density.
Secondly, \citetalias{Hasinger2020} estimates as negligible the heating of accreting gas by X-rays produced by PBHs, which instead in our model provides a negative feedback on the PBH emissivity. Moreover, \citetalias{Hasinger2020} adopts an extended mass function with a peak around $1~\rm M_{\sun}$, but with broad tails reaching up to $10^{9}~\rm M_{\sun}$. They conclude that the dominant contribution arises from PBHs with $M_{\text{PBH}} \sim 10^{4}~\rm M_{\sun}$, while we focused only on the range $1\le M_{\text{PBH}}/\rm M_{\sun} \le 10^3$. A final difference concerns the accretion parameter, which they assume to be $\lambda = 0.05$, i.e. 5$\times$ higher than the one adopted here. 

To summarize, even if our modelling for the PBH contribution to the NIRB excess differs from the \citetalias{Hasinger2020} one, we don't end up with a dramatic discrepancy. This is because the extra physical effects that we have included tend to balance each other. In fact, we should have expected a much higher NIRB flux due to the contribution of PBHs accreting in haloes and the boosted matter power spectrum due to the presence of the PBHs. However, these effects are balanced by the inclusion of the IGM heating from PBH X-ray emission that damps their emissivity at lower redshifts.

\section*{Acknowledgements}
Plots in this paper produced with the MATPLOTLIB \citep{Matplotlib} package for PYTHON.
\section*{Data Availability}
The data underlying this article will be shared on reasonable request to the corresponding author.



\bibliographystyle{mnras}
\bibliography{bibliography} 





\bsp	
\label{lastpage}
\end{document}